# Transport properties and the anisotropy of $Ba_{1-x}K_xFe_2As_2$ single crystals in normal and superconducting states.

V.N.Zverev[1], A.V.Korobenko[1,2], G.L.Sun[3], D.L.Sun[3],
C.T.Lin[3], and A.V.Boris[3,4]

[1] *Institute of Solid State Physics, Russian Academy of Sciences, Chernogolovka, Moscow region, 142432 Russia*
[2] *Moscow Institute of Physics and Technology, Dolgoprudny, Moscow Region, Russia*
[3] *Max-Planck-Institut für Festkörperforschung, Heisenbergstraße 1, 70569 Stuttgart, Germany*
[4] *Department of Physics, Loughborough University, Loughborough, LE11 3TU, United Kingdom*

The transport and superconducting properties of $Ba_{1-x}K_xFe_2As_2$ single crystals with $T_c \approx$ 31 K were studied. Both in-plane and out-of plane resistivity was measured by modified Montgomery method. The in-plane resistivity for all studied samples, obtained in the course of the same synthesis, is almost the same, unlike to the out-of plane resistivity, which differ considerably. We have found that the resistivity anisotropy $\gamma=\rho_c/\rho_{ab}$ is almost temperature independent and lies in the range 10-30 for different samples. This, probably, indicates on the extrinsic nature of high out-of-plane resistivity, which may appear due to the presence of the flat defects along Fe-As layers in the samples. This statement is supported by comparatively small effective mass anisotropy, obtained from the upper critical field measurements, and from the observation of the so-called "Friedel transition", which indicates on the existence of some disorder in the samples in c-direction.

After the discovery of high-temperature superconductivity in the iron arsenides [1,2], both experimental and theoretical activity were directed on the study of the band structure, transport properties and the pairing symmetry in the superconducting state. Despite the intensive studies, many important physical issues concerning the properties of these new materials are still discussed controversially. In particular, this is true for such an important parameter as the anisotropy. The high anisotropy was expected according to band structure calculations [3] and was supported by the experiments in non-superconducting $BaFe_2As_2$ [4], $SrFe_2As_2$ [5] and superconducting electron-doped $BaFe_{2-x}Co_xAs_2$ [6], where the out-of-plane $\rho_c$ to in-plane $\rho_{ab}$ resistivity ratio $\gamma=\rho_c/\rho_{ab}$ was found to be about 100, covering the range between 21 [5] and 150 [4]. Recently the anisotropy was measured in the samples of pristine $AFe_2As_2$ (*A*=Ca, Sr, Ba) [7] and Co-substituted $BaFe_2As_2$ [8] using the Montgomery method and the ratio $\rho_c/\rho_{ab}$ proved to be well below 10. This result is in agreement with the measurements of the upper critical field $H_{c2}(0)$ anisotropy [8], taking into account that this anisotropy has to be equal to about $\gamma^{1/2}$. Such a huge discrepancy in $\rho_c/\rho_{ab}$ values, obtained by different groups, is still unclear. One has to take into account that the anisotropy measurements are often complicated and can contain considerable error when as-grown samples are so thin that the out-of-plane component is hard to measure.

In this paper we have studied the transport properties and the anisotropy of hole-doped superconducting $Ba_{1-x}K_xFe_2As_2$ single crystals with $T_c \approx 31$ K, which, unlike to the parent compounds, do not have the anomalies in $\rho(T)$ dependence due to the structural phase transition. Recent studies have demonstrated that the slightly underdoped $Ba_{1-x}K_xFe_2As_2$ samples preserve microscopically the tetragonal symmetry down to the lowest temperatures, while showing a phase-separated magnetic order below ~ 70 K [9,10].

Single crystals of $Ba_{1-x}K_xFe_2As_2$ were grown using Sn as flux in a zirconia crucible sealed in a quartz ampoule filled with Ar. A mixture of Ba, K, Fe, As, and Sn in a weight ratio of $(Ba_{1-x}K_xFe_2As_2)$:Sn =1:85 was heated in a box furnace up to 850 °C and kept constant for 2–4 hours to soak the sample in a homogeneous melt. An extra of K with 30 wt% was added into the mixtures to compensate the loss from high melting temperature. A cooling rate of 3 °C/h was then applied to decrease the temperature to 550 °C, and the grown crystals were then decanted from the flux. The growth method and the crystal structure and composition characterization are described in detail in Ref. [11]. The samples grown at very the same conditions have been extensively studied by muon-spin rotation [9] and angle-resolved photoemission spectroscopy [12, 13].

Sample resistance was measured using a four-probe technique by a Lock-in detector at 20Hz alternating current in the temperature range (300-4.2) K. We have tested fore samples obtained in the course of the same synthesis. For three of them both in-plane and out-of-plane resistivity tensor components were measured using modified Montgomery method [14]. This method takes into account the real contact positions on the sample surface (see Fig.1) unlike the traditional Montgomery method [15], for which the contacts have to be placed on the corners of rectangular plate. The samples were the plates with about 0.60x0.30x0.15 mm$^3$ characteristic sizes. Two contacts were prepared to each of two opposite sample surfaces, oriented along (ab) plane, with conducting silver paste. In the experiment we could measure either $R_{\parallel} = V_{12}/J_{34}$ or $R_{\perp} = V_{24}/J_{13}$ when the current J was run mainly parallel or perpendicular to (ab) plane respectively (Fig.1). From $R_{\parallel}$ and $R_{\perp}$ values the resistivities $\rho_c$ and $\rho_{ab}$ were calculated. The accuracy of the calculated resistivity values is about 30% and it is determined mainly by the non-ideal shape of the samples. The control measurements were carried out on the thin (about 0.03 mm) sample using standard 4-probe technique. In this experiment in-plane resisitivity tensor component was obtained directly from the sample resistance. On the same sample the Hall measurements and the measurements of the upper critical field were also carried out. According to Hall measurements our samples have p-type conductivity with carrier concentration about $2 \cdot 10^{21}$ cm$^{-3}$.

Typical $R_{\parallel}(T)$ and $R_{\perp}(T)$ dependences are shown in Fig.1 for one of the samples. The results of resistivity measurements $\rho_{ab}(T)$ for all samples are summarized in Fig.2. The curves $\rho_{ab}(T)$ are convex with the tendency to saturate at high temperature that is consistent with the results of the previous reports for hole-doped $Ba_{1-x}K_xFe_2As_2$ [2, 6], whereas $\rho_{ab}(T)$ of electron-doped $BaFe_{2-x}Co_xAs_2$ reveals roughly a linear behavior [8]. The saturation could be brought for the proximity to the so-called Ioffe-Regel limit. At T ≈30 K the mean free path value for our samples $l≈3·10^{-7}$ cm is considerably greater than the lattice parameters ($a$ = $3.9·10^{-8}$ cm, $c$ = $1.3·10^{-7}$ [2]), but $l$ goes down when the temperature increases and near the room temperature these parameters could become comparable. Alternatively, the saturation could be explained in the frame of a two-band model [16]. In the case of two bands with different parameters the conductivity of one band can "shunt" the conductivity of another, leading to the saturation of the total resistance at high temperature. This scenario can also explain the qualitative difference in the shape of $\rho_{ab}(T)$ between electron- and hole-doped systems by a profound difference in their electronic structure [17].

The resistivity anisotropy $\rho_c/\rho_{ab}$, which is almost temperature independent, is presented in the Insert to Fig.2. We would like to emphasize that in-plane resistivity values for all studied samples proved to be close to each other both for Montgomery and 4-probe measurements. This is not true for the out-of-plane resistivity. One can see that the $\rho_c/\rho_{ab}$ values differ considerably for three studied samples and the difference is much higher than the experimental error. In contrast to the dc measurements, the anisotropy ratio extrapolated from our recent far-infrared conductivity measurements is lower by a factor of 2-3 even for the highly conductive #3 sample (blue curve in the inset in Fig. 2) [18]. This result demonstrates that, unlike to $\rho_{ab}$, which is almost the same for all our samples, $\rho_c$ value is considerably differ and, probably, has the extrinsic origin. This phenomenon is well known for the layered systems (graphite, layered semiconductors, etc.) in which the out-of-plane conductivity is limited by the presence of the flat defects.

The superconducting transition temperature $T_c$, determined from $R_{\parallel}(T)$ at the midpoint between 10% and 90% transition level, lies in the interval (29.5-30.5) K for our samples. Interestingly, $T_c$ value proved to be slightly dependent of the current orientation. This effect is demonstrated in Fig.3. As seen from the Figure, for $J\parallel c$ $T_c$ value is about 1K smaller than for $J\parallel (ab)$ . This result does not depend on the current value and, hence, has nothing to do with the electron system overheating which could take place because of the difference in the power dissipation for longitudinal and transversal geometries. The same, but more pronounced effect was observed earlier in the layered high-$T_c$ superconductors [19, 20]. The possible physical reason for the different $T_c$ values obtained from longitudinal and transversal resistance measurements is a layer decoupling transition, the so-called "Friedel transition" [21], which occurs for a disordered layer array [22].

The influence of the magnetic field on the superconducting transition for $B\parallel c$ is shown in Fig.4. One can see that the transition shifts to low temperature region without considerable broadening. For $B\parallel (ab)$ the behavior is similar, but the effect of magnetic field is more feeble. The temperature dependence of the upper critical field $H_{c2}(T)$, obtained from these data, are shown in Fig.5. The slopes $dH_{c2}/dT$ for

B||(ab) and B||*c* near T$_c$ are equal to -12 T/K and -5.0 T/K respectively. Using the Werthammer-Helfand-Hohenberg formula [23] H$_{c2}$(0)=-0.69(d H$_{c2}$/dT|$_{Tc}$ )T$_c$ one gets H$^{ab}_{c2}$(0)=248 kOe and H$^{c}_{c2}$(0)=105.6 kOe for T$_c$=30 K and the critical field anisotropy 2.4. This last value gives the effective mass anisotropy about 5.8. We realize that the effective masses in anisotropic Ginzburg-Landau model are not the same that the masses, which describe the normal state transport properties. Nevertheless, the fact that the resistivity anisotropy is considerably greater than that obtained from the critical field measurements support our statement about the extrinsic origin of the out-of-plane resistivity.

In conclusion, we have measured the anisotropy of transport and superconducting properties of Ba$_{1-x}$K$_x$Fe$_2$As$_2$ single crystals. We have found that the in-plane resistivity for all studied samples, obtained in the course of the same synthesis, is almost the same, unlike to the out-of plane resistivity, which differ considerably. This, probably, indicates on the presence of flat defects parallel to Fe-As layers in the samples. This statement is supported by the comparatively small effective mass anisotropy, obtained from the upper critical field measurements, and from the observation of the so-called "Friedel transition", which indicates on the existence of some disorder in *c*-direction.

**Acknowledgment:** We thank A.A.Golubov and O.V.Dolgov for valuable discussions and suggestions.

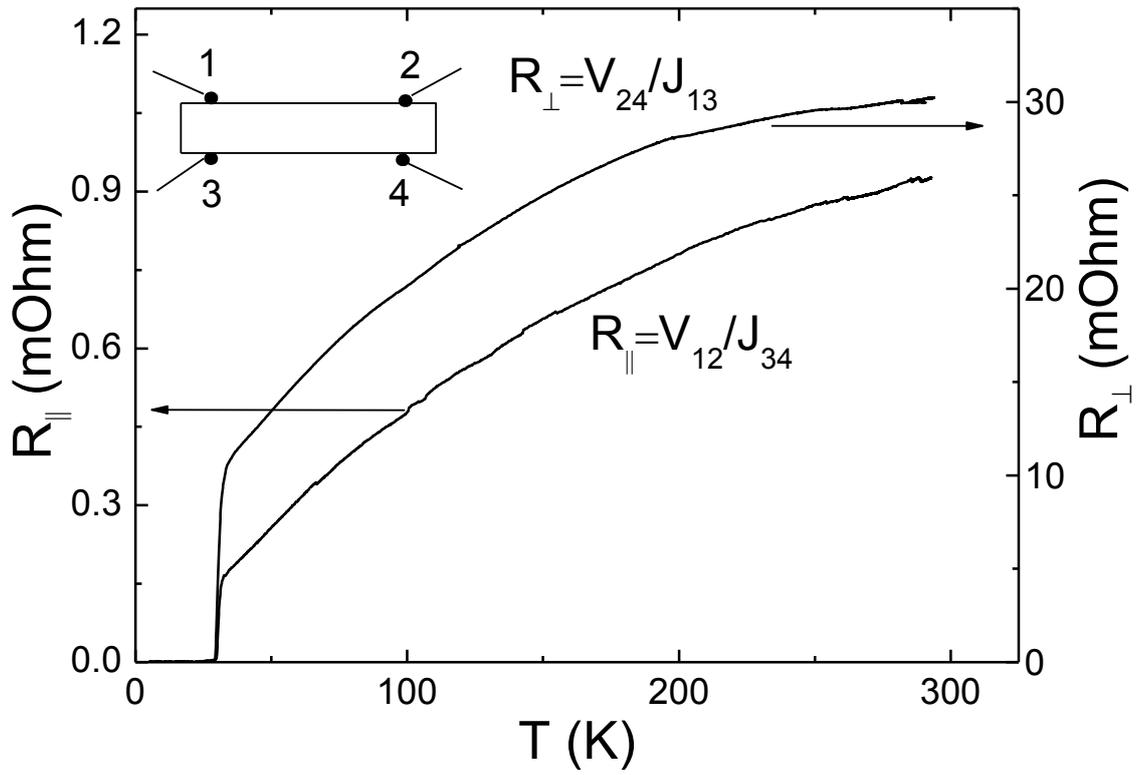

Fig.1. Temperature dependences $R_{\parallel}(T)$ and $R_{\perp}(T)$. The contact positions for Montgomery measurements are shown in the Insert.

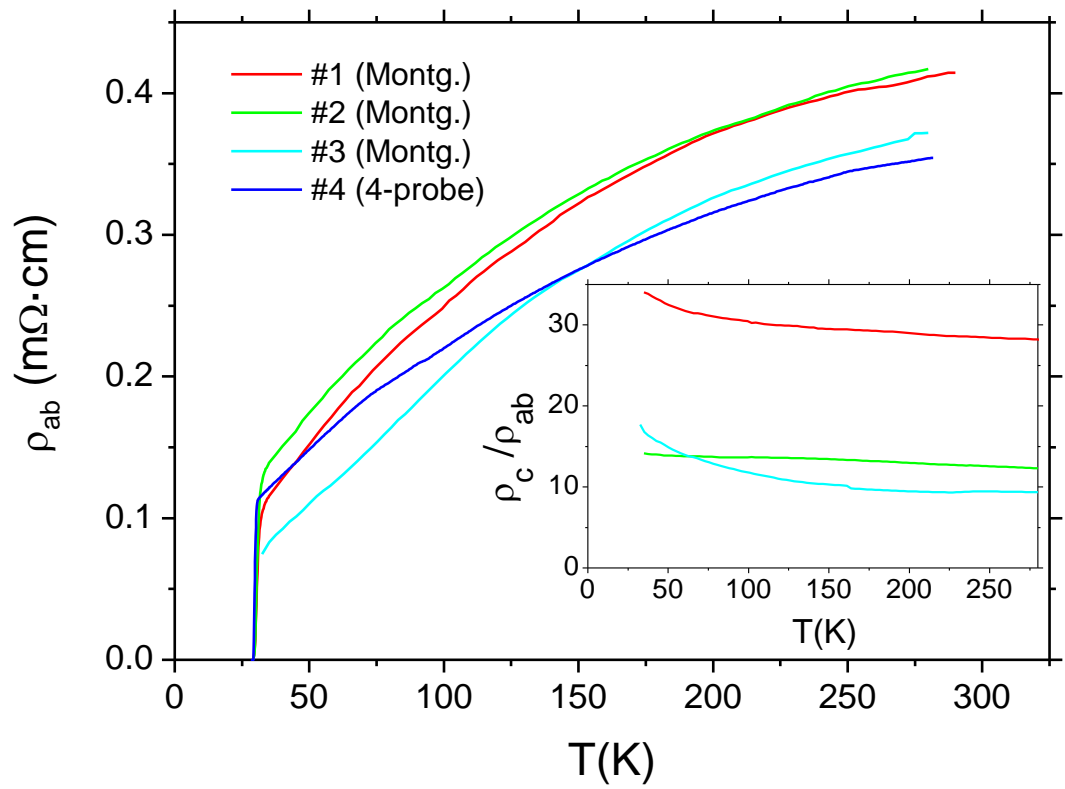

Fig.2. The results of the resistivity $\rho_{ab}(T)$ measurements. The resistivity anisotropy is shown in the Insert.

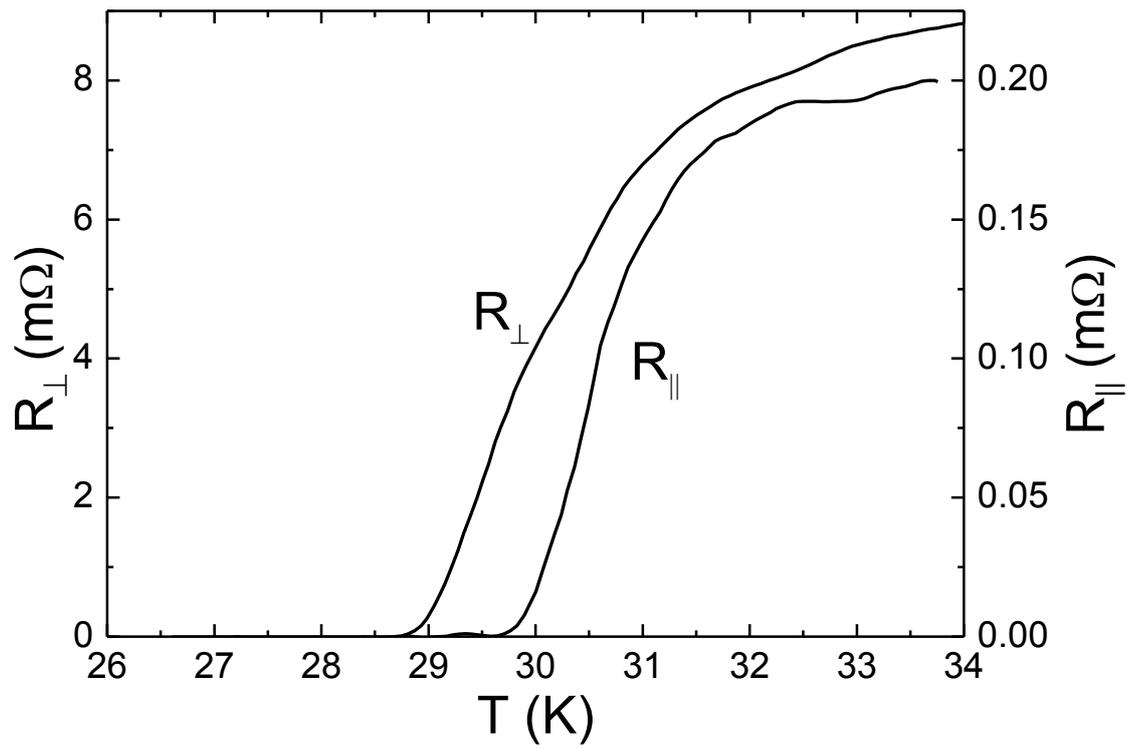

Fig.3. The influence of the current direction on the superconducting transition temperature.

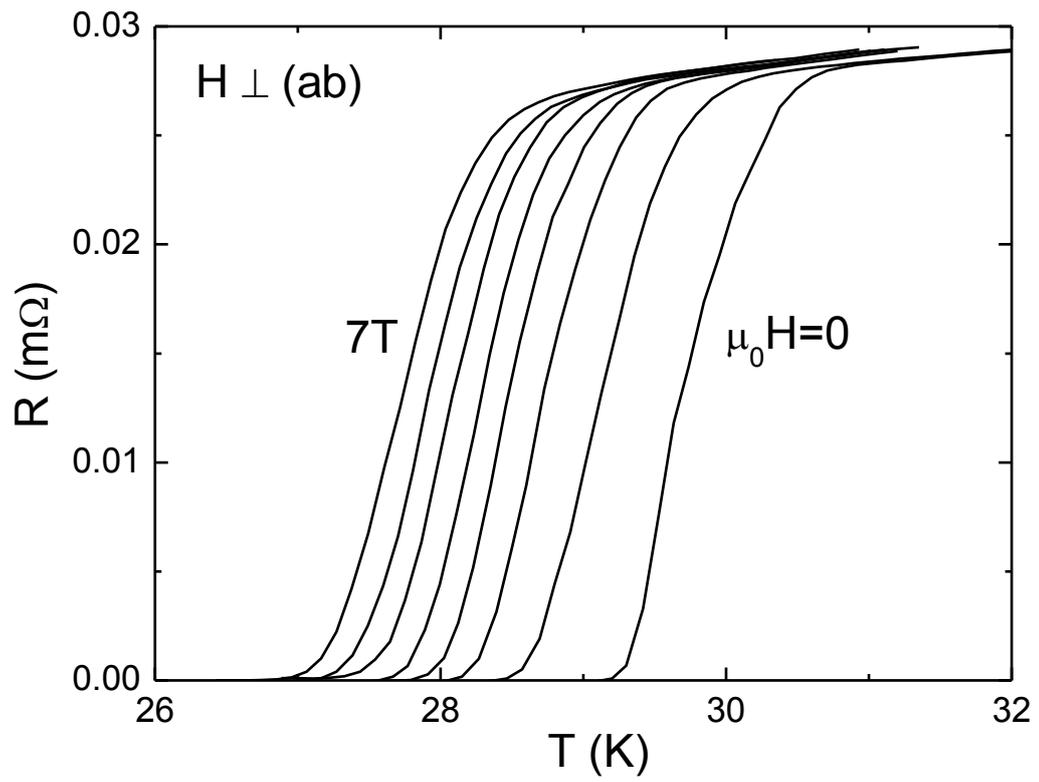

Fig.4. The influence of magnetic field on the superconducting transition for B∥*c*.

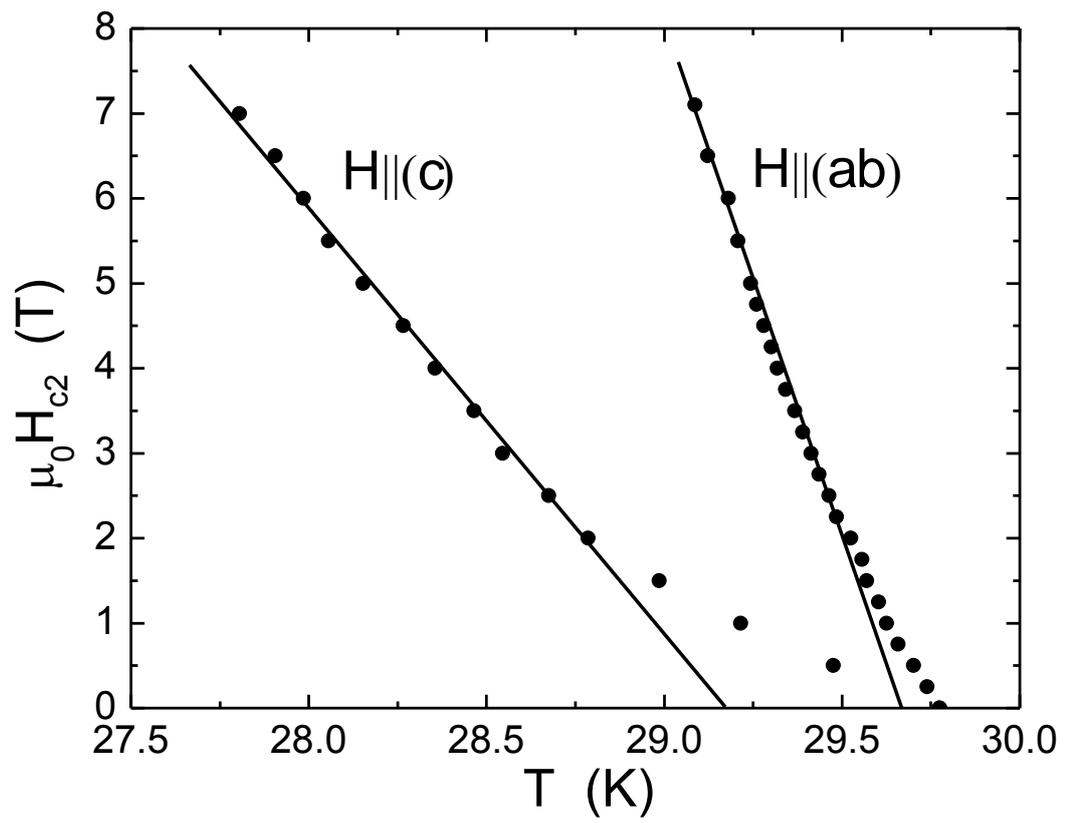

Fig.5. The temperature dependence of the upper critical field $H_{c2}(T)$.